%% file: paper_arxiv.tex
\documentclass[conference,a4paper]{IEEEtran}

\usepackage[subtle,tracking=normal]{savetrees}
\usepackage{tikz}

\usepackage{booktabs}
\usepackage[T1]{fontenc}

\usepackage{algorithm} %For algorithm
\usepackage{algpseudocode} %For algorithm
\usepackage{balance}
\usepackage{amssymb} %Include symbols
\usepackage{graphicx} %For figures

\usepackage{array} %For tables, centering
\usepackage{cleveref}
\crefformat{section}{\S#2#1#3} % see manual of cleveref, section 8.2.1
\crefformat{subsection}{\S#2#1#3}
\crefformat{subsubsection}{\S#2#1#3}
\usepackage{url}

\usepackage{hhline} % For table double line
\usepackage{array}
\newcolumntype{P}[1]{>{\centering\arraybackslash}p{#1}}
\newcolumntype{M}[1]{>{\centering\arraybackslash}m{#1}}

\begin{document}

\title{Providing In-network Support to Coflow Scheduling}

\author{\IEEEauthorblockN{Cristian Hernandez Benet\IEEEauthorrefmark{1},
Andreas J. Kassler\IEEEauthorrefmark{1}, Gianni Antichi 
\IEEEauthorrefmark{2} 
\\
Theophilus A. Benson\IEEEauthorrefmark{3} and Gergely Pongracz\IEEEauthorrefmark{4}}
\\
\IEEEauthorblockA{
\IEEEauthorrefmark{1}Karlstad University,
\IEEEauthorrefmark{2}Queen Mary University,
\IEEEauthorrefmark{3}Brown University,
\IEEEauthorrefmark{4}Ericsson Research}
\\
\IEEEauthorblockA{
\IEEEauthorrefmark{1}{cristian.hernandez-benet,andreas.kassler}@kau.se,
\IEEEauthorrefmark{2}g.antichi@qmul.ac.uk,
\IEEEauthorrefmark{3}tab@cs.brown.edu,
\IEEEauthorrefmark{4}Gergely.Pongracz@ericsson.com}}

\maketitle

\IEEEpubidadjcol

\begin{abstract}
\input{abstract.tex}
\end{abstract}


\IEEEpeerreviewmaketitle

\section{Introduction}
\label{sec:intro}

\input{intro.tex}

\section{Motivation}
\label{sec:Motivation}
\input{motivation}

\section{Design}
\label{sec:implementation}
\input{implementation.tex}

\section{Evaluation and Results}
\label{sec:eval}
\input{eval.tex}

\section{Related Work}
\label{sec:background_motivation}
\input{background_motivation.tex}

\section{Conclusions and Future Work}
\label{sec:conclusion}
\input{conclusion.tex}

\section*{Acknowledgment}
 Parts of this work is
   supported by the Knowledge Foundation of Sweden through the Profile HITS under Grant No.: 20140037.

\bibliographystyle{IEEEtran}

\bibliography{paper_arxiv.bib}

\end{document}

%% file: abstract.tex
Many emerging distributed applications, including  big data analytics, 
generate a number of flows that concurrently transport data across 
data center networks. To improve their performance, it is required to 
account for the behavior of a collection of flows, i.e., coflows, 
rather than individual. State-of-the-art solutions allow for a near-optimal 
completion time by continuously reordering the unfinished coflows at 
the end-host, using network priorities. 

This paper shows that dynamically changing flow priorities at the 
end host, without taking into account in-flight packets, can 
cause high-degrees of packet re-ordering, thus imposing pressure
on the congestion control and potentially harming network performance
in the presence of switches with shallow buffers. We present \textit{pCoflow},
a new solution that integrates end-host based coflow ordering with 
in-network scheduling based on packet history. Our evaluation 
shows that \textit{pCoflow} improves in CCT upon state-of-the-art solutions by up to 34\% for varying load.

%% file: intro.tex
Emerging big data processing frameworks such as MapReduce~\cite{dean2008mapreduce}, 
Spark~\cite{zaharia2010spark} or Dyan~\cite{isard2007dryad} are based on a partition/aggregate 
programming model that allow them to distribute and parallelize the processing across different 
machines. Such big data analytics frameworks are also becoming an important enabler for future mobile networks with use cases ranging from incident detection at Network Operations Centers~\cite{NOC2018}, Traffic Classification and Network Slicing~\cite{Le2018} to IoT data processing~\cite{Nejkovic2019}. A common property of such big data frameworks is that each processing stage cannot complete until all data have been transferred. As a consequence of this property, the performance of these frameworks is function of the behavior of the 
collection of flows used to transfer data in each stage~\cite{isard2007dryad,zaharia2012resilient}, i.e., coflows~\cite{chowdhury2012coflow}. More formally, coflows are collection of flows of varying sizes with different communication endpoints.

Most of the growing work on improving performance within data centers build upon the advances in data center load balancing techniques, e.g., Hula~\cite{katta2016hula}, and data center centric transports, e.g., DCTCP, which improve low level details of the data center's network. By abstracting details about load balancing, routing, and transport, these emerging techniques can focus on the crucial aspect of the network which impact individual flow performance, i.e., controlling queuing, priorities, or rate-limits. However, existing approaches for queuing, priorities, rate-limits within the data center do not provide the levels of dynamicity required to support recent coflow proposals, e.g., Sincronia~\cite{Sincronia}.

In this work, we ask the following fundamental question: \textit{``Does the network provide sufficient primitives to faithfully support dynamic modification of coflow priorities and queues?''}.
To answer this question, we use a large scale trace-driven simulation, which allows us to methodologically analyze a broad range of existing techniques and scenarios. 
Our initial observations are that current network primitives do not effectively support arbitrary modifications and changes to a coflow's queuing priorities. In particular, we observe that while the network provides the illusion that all packets in a flow can be atomically moved between queues. In practice, once a packet has been queued, it does not dynamically change queues, which leads to packets from a flow ending up in different queues. The end result of this phenomenon is that packets from a flow are spread across queues resulting in a high degree of packet reordering when they arrive at the destination end-hosts which lead to reduced performance due to TCP's design. 

In particular, due to TCP's behavior high-degrees of packet re-ordering can in some cases cause the congestion control window to shrink with negative effect on performance.

In this paper, we argue that existing data center networks lack in-network support for dynamically changing coflow queuing priority: specifically, with existing network primitives changing a flow's priority does not consider packets already traversing the network, thus causing inconsistencies between 
the in-flights and newly generated packets of the flow at switches with multiple priority queues.
Motivated by these observations, we propose an in-network primitive, called \textit{pCoFlow}, which allows flows to temporarily maintain queue affinity until already enqueued packets are drained when flows are being reprioritized due to coflow order update.
A key challenge in preserving flow affinity under re-prioritization lies in maintaining network state and using this state to dynamically override packet priorities and alter queuing behavior.
Our work builds on emerging techniques for programmable data planes~\cite{bosshart14,sivaraman2016programmable} and uses them to maintain minimal per-coflow state, i.e., optimized packet histories, and dynamically manage flow priorities and queue assignments based on this state.

To demonstrate the strength of our approach, we propose the design (\S\ref{sec:implementation}) of a system that integrates state-of-the-art 
ordering mechanisms at the end-host, such as Sincronia, with in-network scheduling 
based on packet history (i.e., \textit{pCoFlow}). We then show that the latter can be implemented in P4 starting 
from the PIFO abstraction~\cite{sivaraman2016programmable}. Finally, we demonstrate
that our approach reduces the average CCT by 15\% up to 18\% for 10\% load and by 27 up to 34\% for 90\% load with respect to state-of-the-art
solutions. 

In summary, the contributions of this paper are:
\begin{itemize}
\item We make the case for in-network support in the context of coflow scheduling.
\item We propose \textit{pCoflow}, which is an architecture that integrates state-of-the-art techniques performing ordering at the end-host with advanced in-network packet scheduling.
\item We design a coflow aware in-network priority scheduler that avoids reordering and can be implemented using the PIFO abstraction~\cite{sivaraman2016programmable}.
\end{itemize}

%% file: motivation.tex
Recently, there has been a tremendous effort on network designs for coflows~\cite{zhao2015rapier,Chowdhury_2014,li2016efficient,chowdhury2011managing,luo2015minimizing,susanto2016stream}.
Some of the works advocate for adopting a distributed approach~\cite{luo2015minimizing,susanto2016stream},
where coflows are scheduled and managed locally at the end hosts; others propose a centralized 
scheme~\cite{zhao2015rapier,Chowdhury_2014,li2016efficient,chowdhury2011managing}, where a 
single entity  with global knowledge is in charge of managing coflows. Centralized 
solutions that  rely on global knowledge have proved to guarantee better performances.
However, the need to centrally calculate complex per-flow rate allocations has hindered the 
possibility to realize them in practice~\cite{Sincronia}. 

Recently, Sincronia~\cite{Sincronia} has demonstrated that the key ingredient to obtain near-optimal performances is to provide convenient coflow  prioritization. Specifically, 
if the \textit{right} ordering of coflows is given, any per-flow rate allocation mechanism would 
lead to \textit{good} results provable close to the optimum if co(flow) scheduling preserves the order. In other words, if coflow $C_i$ is ordered higher than coflow $C_j$, all flows and packets in $C_i$ must be prioritized over $C_j$.
Such an important finding has opened up the possibility of a 
new network design where a central controller is just in charge of ordering the coflows, while 
leaving any per-flow prioritization to the end-hosts, thus striking the right balance between 
centralized and distributed schemes. In this regard, Sincronia assumes that individual flow 
scheduling and rate allocation is provided by a priority-enabled transport layer at the end-host, 
obeying to a centrally managed coflow ordering controller~\cite{Sincronia}. In practice, given 
a coflow ordering, the end-host will continuously (re)assign the priority of a flow coherently 
with the coflow it belongs to using, for example, DiffServ markings.

\begin{figure}[!htb]
    \begin{minipage}[t]{0.23\textwidth}
     \centering
     \includegraphics[width=.999\linewidth]{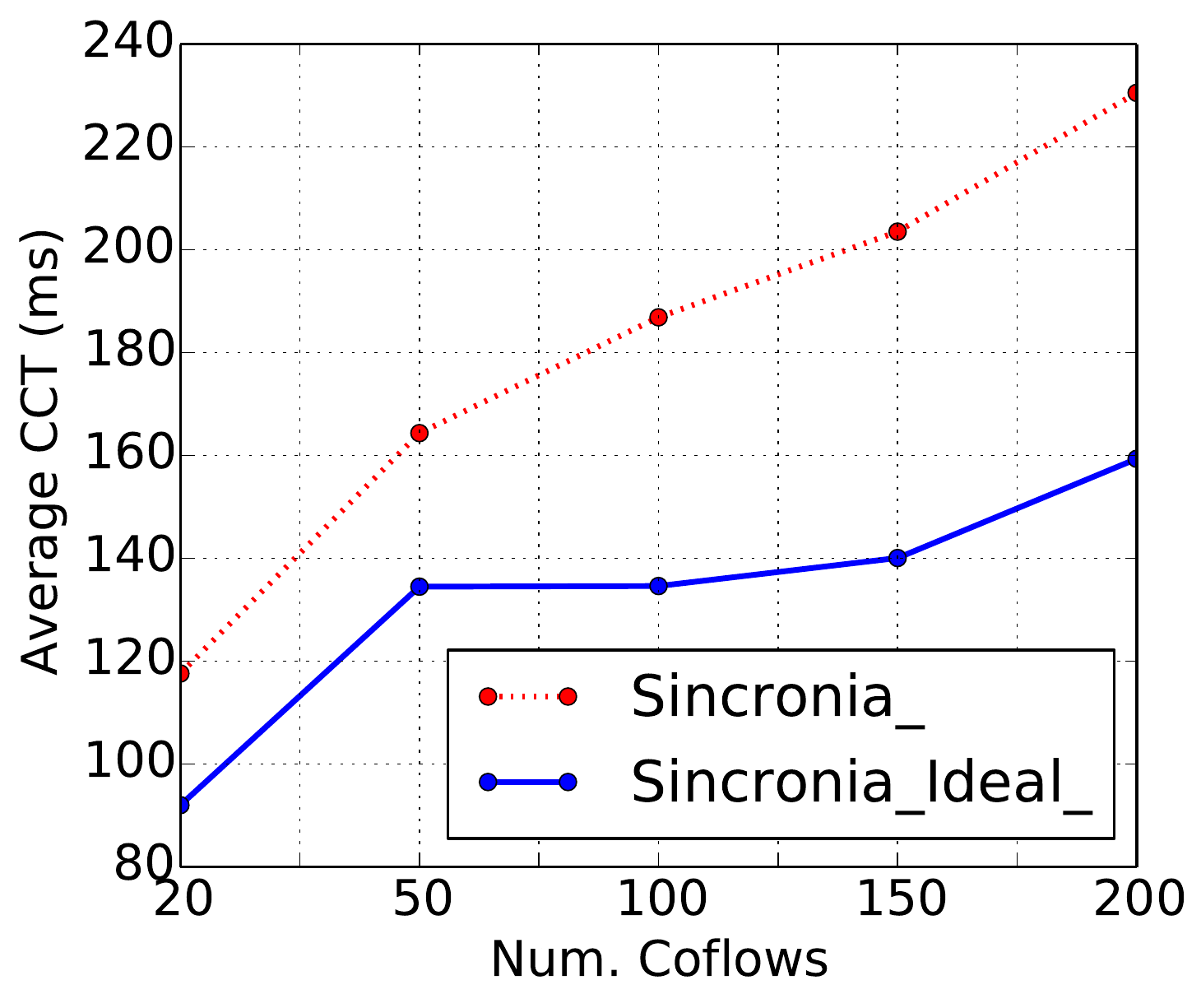}\par\caption{Average coflow completion time} \label{fig:CCT_motivation1}
   \end{minipage} \hfill
    \begin{minipage}[t]{0.23\textwidth}
     \centering
     \includegraphics[width=.999\linewidth]{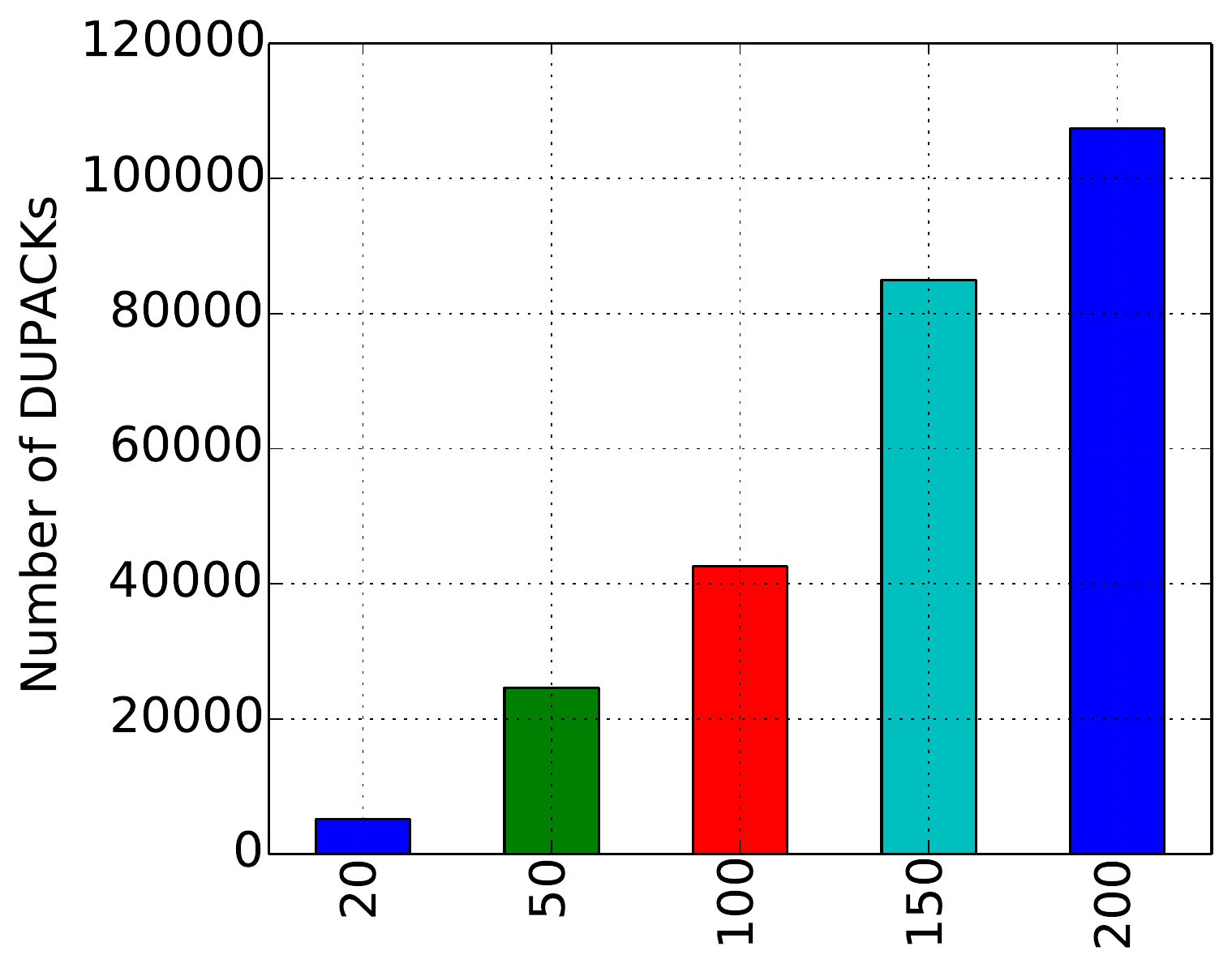}\par\caption{DupAcks for different number of coflows} \label{fig:DupAcks_Motivation}
   \end{minipage} \hfill
\end{figure}

\noindent\textbf{The need for in-network support.} Decoupling scheduling decisions, centrally 
managed, from the flow-rate allocation problem, controlled at the end-hosts through a priority-enabled transport layer, might generate 
an excessive amount of out-of-order packets thus affecting network performances. To illustrate 
this problem, we used the NS2 simulator. We assumed a non-blocking big-switch as network 
topology~\cite{alizadeh2013pfabric,Chowdhury_2014,Sincronia} 
and we used Data center TCP (DCTCP)~\cite{alizadeh2010data} as state-of-the-art congestion control for data center 
networks. We generated traffic according to the Sincronia workload generator~\cite{Sincronia} 
which is based on Facebook traffic characteristics and used an increasing number of coflows 
from 20 to 200. We let Sincronia calculate the coflow ordering and we enforced the corresponding 
flow priority using DSCP marking. Furthermore, to properly enforce 
the correct ordering, we enhanced the big-switch abstraction with eight queues per port.
Figure~\ref{fig:DupAcks_Motivation} shows the total number of timeout events obtained for 
different coflow sizes. The reason lies in the Sincronia behavior that dynamically change 
flow priorities enforcing new policies from the end hosts without considering packets that 
are already traversing the network. Duplicated ACKs might trigger a flow to shrink its congestion 
window, impacting directly on network performance.
The effect on the CCT is shown in Figure~\ref{fig:CCT_motivation1}.
Here, we compare the CCT we obtained with Sincronia against an optimal scenario where a change 
in coflow priority does not cause any packet reordering. The ideal case performs up to 
1.5x better than Sincronia. 

\begin{figure}[!htbp]
  \centering
   \vspace{-1mm}
  \includegraphics[width=0.45\textwidth,keepaspectratio]{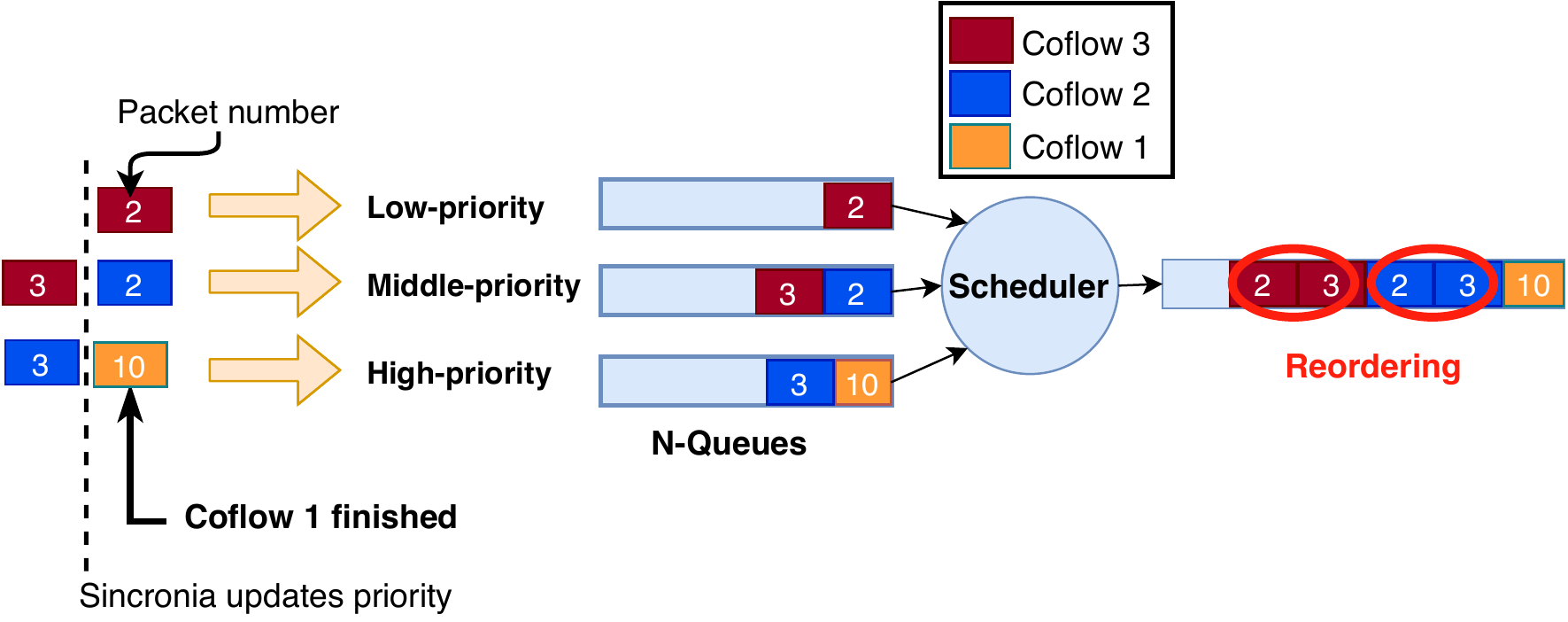}
  \caption{Packet reordering after priority updates}
  \label{fig:CCT_motivation}
  \vspace{-1mm}
\end{figure}

To better understand the cause of packet reordering, we illustrate in Figure~\ref{fig:CCT_motivation}
what happens when the Sincronia controller issues a change in coflow priority.
Let us assume that at some point coflow 1
finishes and Sincronia increases the priority level of each remaining active
coflow. This change affects new packets to be generated from end hosts, while 
the one already in-flight will be served with the old configuration, i.e., 
priority. This clearly creates packet reordering if packets of the same flow are still enqueued at a lower priority queue at the same switch and newer packets having higher priority arrive due to strict priority queue. Reordering may trigger congestion control, which reduces the rate of the flow.

This simple example illustrates a wider 
problem: in real data center topologies where multiple paths from source to 
destination are available, the amount of reordering as well as its effect on 
network performances can be even bigger. Indeed, to select the \textit{best path},
the research community has shown the effectiveness of congestion aware flowlet-based 
load-balancing approaches~\cite{alizadeh2014conga,katta2016hula}. Those schemes 
split flows into smaller \textit{flowlets}, exploiting the burstiness of TCP. The
idea is to route each flowlet over the least congested path. However, when using
Sincronia in combination with the mentioned solutions might trigger a reordering
of not \textit{just few packets}, but instead \textit{entire flowlets}.

%% file: implementation.tex
Given the insights from the previous section, we ask the question \textit{is it possible 
to minimise packet reordering due to priority changes by allowing switches to participate 
in scheduling decisions?} We answer this positively by describing \textit{pCoflow}, a solution 
which provides in-network support for coflow scheduling. \textit{pCoflow} integrates 
state-of-the-art ordering techniques at the end-host, e.g., Sincronia, with scheduling 
decisions taken \textit{in-network}. The main idea is to leverage programmable switches to temporarily maintain queue affinity for newly arriving packets  until already enqueued packets are drained when flows are being reprioritized due to coflow order update. We show that this can be realized with the PIFO abstraction~\cite{sivaraman2016programmable} and by taking into 
account the priorities of packets \textit{before} and \textit{after} an update, i.e., their history.

\subsection{Design Objectives} \label{Objectives}
\noindent
\textbf{Avoiding In-Network Reordering:}

Coflow schedulers such as Sincronia ensure coflow isolation and preserve the order of coflows by delegating prioritization to a priority-enabled transport mechanism. When run together with TCP, this requires a prioritization of IP packets according to the coflow priority, which is typically implemented using a multi-level queuing system with strict priorities. Such systems schedule packets waiting at higher priority queue first. Ideally, this would allow higher priority coflows to finish earlier and thus improving overall transmission time. However, the arrival, termination or changes in the remaining transmission time of coflows can lead to shifting in priority levels at end-hots that might lead to packet reordering, triggering congestion control and reducing the rate of the newly prioritized flow, achieving exactly the opposite effect. When using multi-level queuing systems, priority updates at end-hosts should therefore not result in packet reordering. Therefore, we aim to implement a single queue that manages coflow priorities during insert using packet histories that track priorities of enqueued packets.

\noindent
\textbf{Avoiding Coflow Starvation:} \label{starvation}
Whether coflows are prioritized using per-flow rate allocation or transport layer priorities, coflow starvation must be avoided. This is necessary to ensure that large coflows do not starve short coflows.  By leveraging network feedback in the form of Early Congestion Notification (ECN), congested network elements can signal end-host transport layer such as TCP~\cite{floyd1994tcp} or DCTCP~\cite{alizadeh2010data} that congestion is building up forcing them to scale back in rate. However, when using a single queue that manages different priorities, care must be taken to how ECN marking is applied. 

\subsection{\textit{pCoflow} Design Overview} \label{overview_design}
\begin{figure}[!htbp]
  \centering
   \vspace{-1mm}
  \includegraphics[width=0.45\textwidth,keepaspectratio]{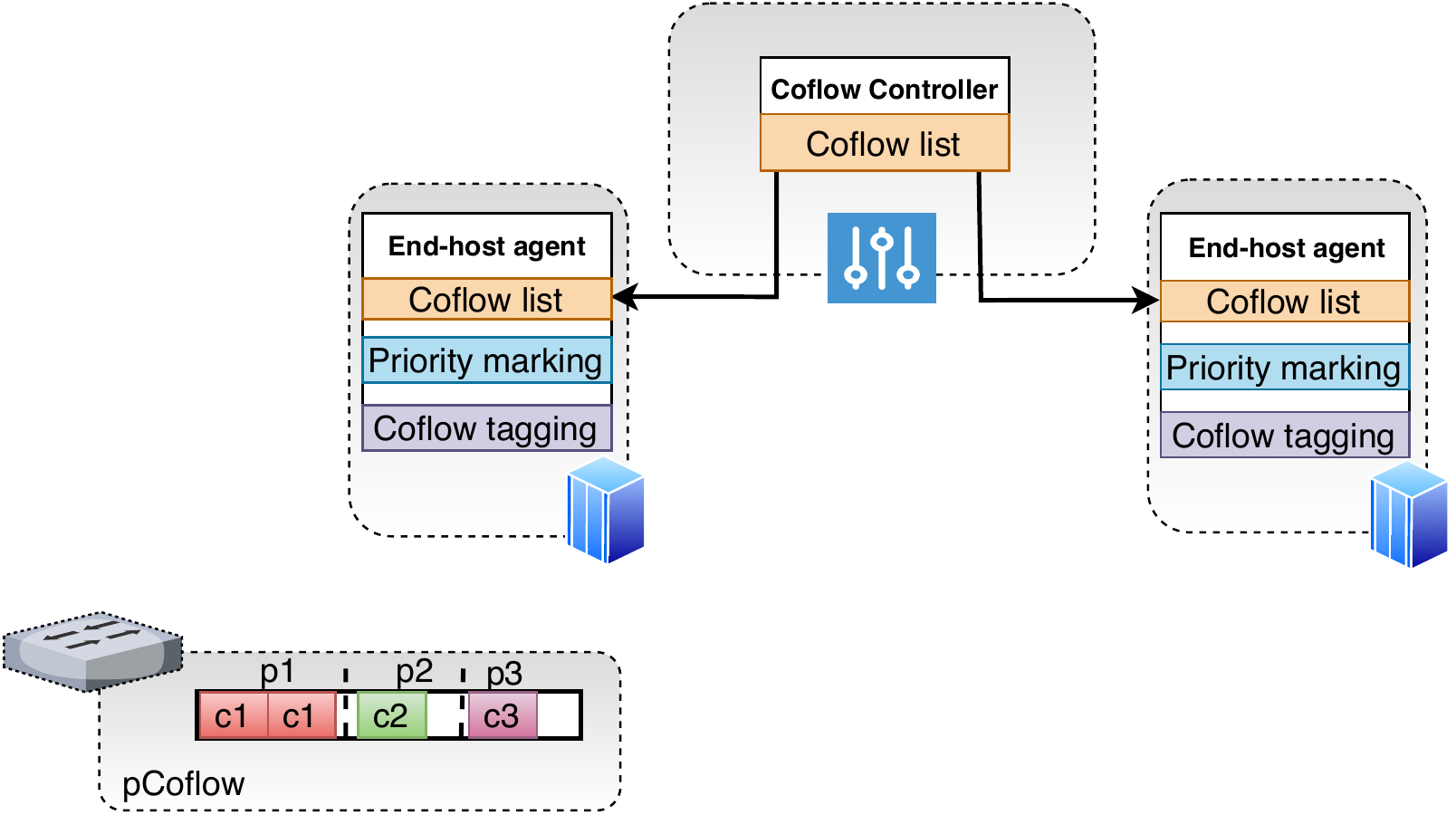}
  \caption{pCoflow Architecture Overview}
  \label{fig:pCoflowArchitecture}
  \vspace{-1mm}
\end{figure}

\textit{pCoflow} uses state-of-the-art centralized coflow controllers such as Sincronia that orders coflows and derives their priority and combines it with transport layer that enforces priority scheduling (see Figure  \ref{fig:pCoflowArchitecture}). The core component of \textit{pCoflow} is a novel coflow aware strict-priority packet scheduler inside the data plane that is aware of priority levels of coflow packets waiting in the queues. 

\subsection{End-host} \label{endhost}
End-hosts are responsible for marking packets with the corresponding coflow priority and sending packets over the transport protocol. We exploit a shim layer between the application and the transport layer which continuously orders the coflows using e.g. information available from a centralized controller (e.g. Sincronia)~\cite{Sincronia}). The coflow order is translated by the end-host agent to DSCP values that map the highest order coflow to the highest priority level. Second highest priority is mapped to second highest priority, etc. and all remaining priorities are mapped to the lowest priority level. The shim layer also tags each packet with an unique \texttt{coflowID}, which is subsequently used by switches to avoid reordering when coflow order is updated by the end-host shim layer. The \texttt{coflowID} can be provided in an extra header (e.g. using GPE extension of VXLAN) or can be conveyed within the IP Identification field or TCP options.

\subsection{In-Network Coflow-aware Scheduler} \label{data_plane}
The main objective of our coflow aware programmable scheduler approach is to maintain coflow priorities for dequeuing operation while avoiding reordering when coflow priorities are switched at end-hosts. Consequently, we need to maintain the relative scheduling order of buffered packets with future packet arrivals, which will be implemented during the push-in operation. The main idea of our approach is to assigning coflows to a single queue as long as buffer space is available. In order to enable prioritization of coflows, we therefore partition a single queue into multiple virtual priority bands, each one having a dedicated priority level and a certain buffer space. As long as there is space available at a given priority level, coflow packets matching that priority level can be inserted appropriately.

Packet reordering may only happen when the end-host increases the priority level of a given coflow. If there are still packets enqueued at the switch for the same flow 
at lower priority levels, the newly arriving packets are served first, leading to reordering. On the contrary, when new packets have a lower priority, there will be no reordering as packets waiting at higher priority levels will be served first. In order to avoid packet reordering due to end-host triggered coflow order update, we first identify the coflow priority by parsing the DSCP field in the packet header

The insert operation has to ensure that packets with highest priority are inserted at the first priority band of the queue, and packets with lower priority at lower priority bands. This makes sure that packets with higher priority are always served first implementing a strict priority queuing policy. In order to avoid that a change in coflow priorities may lead to packet reordering within a flow, we need to check to which coflow a packet belongs to. If a Sincronia triggered reorder increases the priority of coflow $C_j$, a packet may arrive at a switch with higher priority and several packets of the same coflow $C_j$ may wait for transmission at a lower priority level. In this case, we temporarily do not use the higher priority as indicated by the end-host reordering but rather insert the newly arriving packet after packets of the same coflow $C_j$. This avoids reordering and makes transport layer transparent to priority changes. The drawback is a delayed response to priority changes in the switch.

Our packet scheduler needs to store (1) the bounds of priority bands, and (2) for each coflow the lowest priority band that has packets waiting to be served (Figure \ref{fig:PIFO_Coflow}. When a packet with priority $p_i$ that belongs to coflow $C_j$ reaches the switch, the scheduler thus checks the position of the last packet enqueued at priority level $p_i$ and the position of the last packet enqueued for coflow $C_j$ and calculates the rank of the packet as in Equation \ref{max_pos}. 
\begin{equation}  \label{max_pos}
 rank = max(p_i,C_j)+1
 \end{equation}

\textit{pCoflow} uses ECN to signal congestion to the end-host transport layer~\cite{wu2012tuning,bai2016enabling}. When using a single queue with multiple priority bands and a single ECN marking threshold may lead to marking mostly lower priority packets which may lead to coflow starvation. To prevent this effect, \textit{pCoflow} uses multiple ECN marking thresholds, one per priority band. If during enqueue we detect that the number of packets enqueued for a given priority $p_i$ is larger than the ECN threshold for the given band, we set the ECN bit. 

Although the minimum and maximum marking threshold of each priority level can be adjusted to react earlier to congestion and start marking packets before reaching the maximum threshold~\cite{zhu2015congestion}, congestion control algorithms can take several RTTs to adjust the sending rate after receiving ECN notification. Therefore, queue sizes may temporarily exceed the defined ECN thresholds. Dropping packets may be necessary when using transport protocols that do not react to ECN or if we want to guarantee a certain buffer space for each priority. However, enforcing packet drop reduces queue elasticity. \textit{pCoflow} enables adaptive queue sizes by dynamically allowing priority bands to increase and shrink. It integrates such dynamic resizing with ECN marking to signal end-points to reduce their rate. 

\subsection{Implementation Feasibility} \label{Feasibility}
With today's switches, only a limited number of scheduling approaches is available whose parameters can be controlled by network operators. However, using programmable packet schedulers allows to implement custom scheduling disciplines tuned to application requirements. Indeed, using the push-in first-out (PIFO) scheduling abstraction~\cite{sivaraman2016programmable}, where packets can be pushed into an arbitrary position but always dequeued from the head, several scheduling approaches can be implemented on programmable data-planes such as P4~\cite{bosshart14}.

\begin{figure}[!htbp]
  \centering
   \vspace{-1mm}
  \includegraphics[width=0.37\textwidth,keepaspectratio]{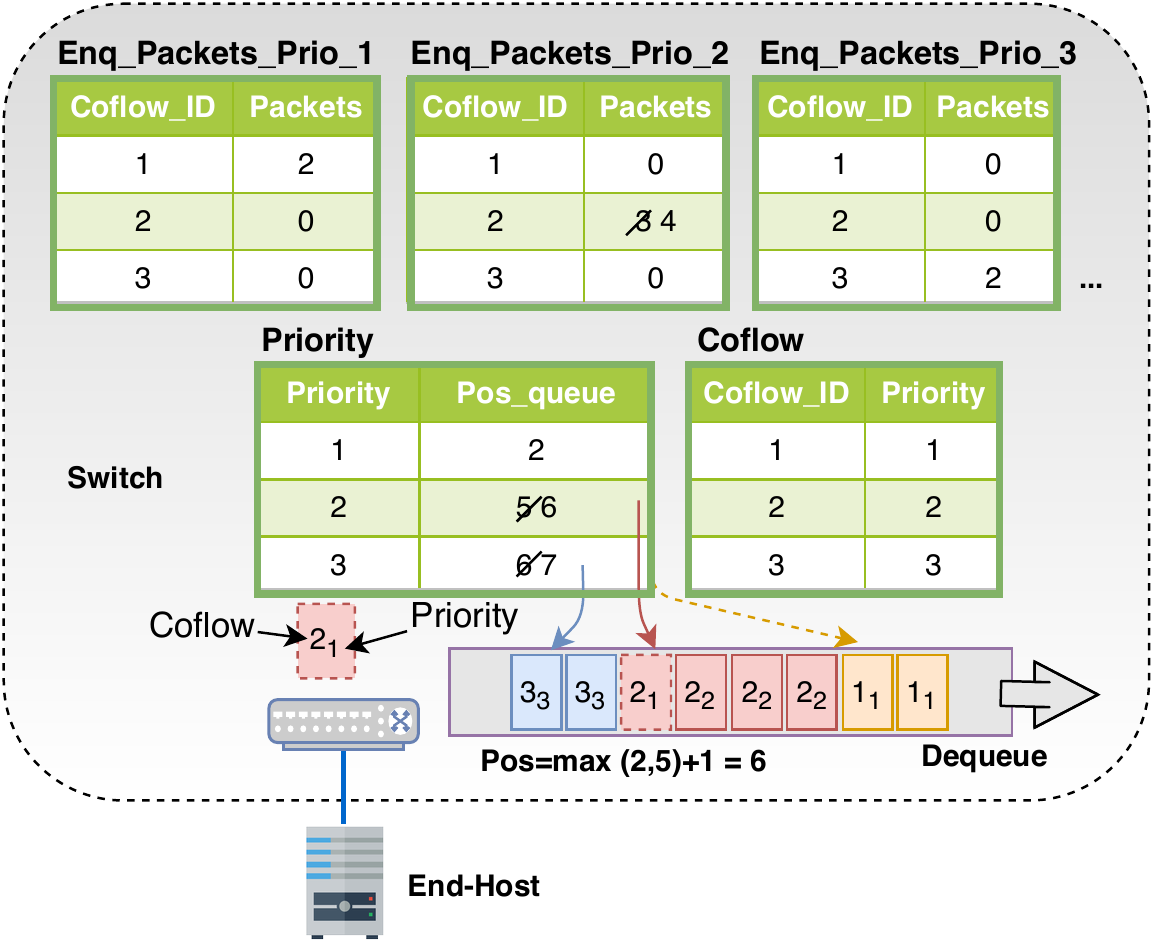}
  \caption{pCoflow Queue with coflow and priority tracking}
  \label{fig:PIFO_Coflow}
  \vspace{-1mm}
\end{figure}

To implement our approach, we can leverage the PIFO abstraction~\cite{sivaraman2016programmable}, which allows enqueued packets to be pushed in arbitrary positions, given by the packet rank, while being dequeued from the head. PIFO assumes that packet ranks increase continuously within a flow and dynamic reordering of already enqueued packets is not supported. Consequently, we need to consider three main issues (i) extracting the coflow priority from the packet header; (ii) mapping packets to correct priority bands and computing the rank; and (iii) updating priority band bounds. For (i), we read the priority bits from the IP header ToS field. Figure \ref{fig:PIFO_Coflow} illustrates our approach.

\noindent
\textbf{Mapping:}
We use registers \texttt{Priority} to store the bound information for each priority band. Assuming $p$ priority bands, we use a register to encode the end of priority band $p_i$. For each coflow, we track the lowest priority band that still has packets  enqueued in register array \texttt{Coflow}. If there is no packet enqueued for a coflow, we set the priority to 0.  
For calculating the PIFO rank at insert and avoid reordering, we first check \texttt{Coflow} to determine, which lowest priority band has packets enqueued (e.g. 2 in the example). Then, we look up the position of the last packet in this band using \texttt{Priority}, which returns 5. We compare this with the rank of the last packet in the priority band that corresponds to the priority marked in the packet header (as the priority is one, the lookup returns 2). Equation \ref{max_pos} returns $rank=max(2,5)+1=6$ which is used for PIFO insert operation.

\noindent
\textbf{Update:} 
As in \cite{Sincronia}, we aim to map each coflow to the given priority band if the current order of the coflow is less than $p-1$, else we map it to band $p$. To avoid reordering when packets are enqueued at priority $p_i$ and new packets for the same coflow arrive having higher priority, $p_{i-k}$, we track which bands have packets enqueued for each coflow using one register array \texttt{Enq\_Packets} for each priority band. We update \texttt{Coflow} as follows. On enqueue of a packet at the end of priority band $p_i$, we update priority band bounds of all lower priority bands $p_{i+k}$ (e.g. if $p_i==2$ we will update bounds of bands 2, 3, 4...). We update \texttt{Enq\_Packets} accordingly (e.g. indicating that priority band 2 has now 4 packets waiting for coflow 2). If there is no packet waiting to be transmitted in any queue (\texttt{Coflow} returned 0), we update \texttt{Enq\_Packets} using the priority band corresponding to the packet priority. On dequeue, we update \texttt{Enq\_Packets} for the priority band we dequeued from and coflow id. If after the dequeue there are no more packets in the current priority band, we sweep the remaining lower priority bands to find the lowest priority band that has still packets waiting and update \texttt{Coflow} for the given coflow id. If there are no more packets waiting, we set \texttt{Coflow} to zero. Finally, we also update \texttt{Priority} of the priority band corresponding to the packet that has been dequeued and all lower priority bands. To track the ECN marking threshold, we use counters per priority band. If we detect that an insert leads to more packets than allowed according to the ECN threshold for a given band, we mark the ECN bit. In our example (Figure  \ref{fig:PIFO_Coflow}), if the ECN threshold is set to 2 packets, when the packet belonging to coflow 2 and priority 1 arrives at the switch, the counter associated to priority 1 will return 2, and the ECN bit is set.  

\noindent
\textbf{Remarks:} 
Note, \textit{pCoflow} downprioritizes temporarily all coflow packets if there are other packets waiting at lower priorities, which maybe not necessary. However, a more fine-granular per flow decision would require per flow tracking, which may lead to excessive switch resources. Note, that all sweeping operations through the multiple priority bands can be implemented as nested \texttt{if-else} statements as the number of bands is determined at compile time. As \textit{pCoflow} does not require to maintain per-flow state (just per priority band and coflow), the required state variables is reasonably small. Increasing the priority bands \emph{p} leads to more fine granular prioritization but requires more switch resources.  Note, the scheme cannot be implemented on state-of-the-art hardware such as Tofino because registers in egress is not available in ingress, which however could be solved by packet recirculation at the expense of higher complexity. PIFO on the other hand supports not more than around 1000 flows~\cite{sivaraman2016programmable}. A variant of our scheme supporting a fixed-size priority bands with limited reordering could be implemented using SP-PIFO~\cite{SP-PIFO}.

%% file: eval.tex
We implemented our scheme in the NS2 simulator and used coflow traces for comparing our scheme against different configurations.
\noindent
\textbf{Topology:} We use a 3-tier Fat-tree topology with k=4. All links have a capacity of 40Gbps, except links connecting 64 servers to the ToRs, which have a capacity of 10Gbps. Each of the 8 ToRs has 8 servers connected that send coflows according to a given trace. Servers run a client application with the Sincronia shim layer that informs the Sincronia coordinator about the coflow information such as coflow id, number of flows and sources, and destination for each flow. It receives the coflow ordering and tags coflow priorities and coflow IDs.
\noindent
\textbf{Coflow Scheduler:} We use the online Sincronia algorithm from \cite{Sincronia} to order coflows. We immediately recompute the order upon each coflow arrival and departure. As in \cite{Sincronia}, we map coflow order to the Diffserv option and use 8 priority levels.

\noindent
\textbf{Workload:} We use~\cite{Sincronia_loadgenerator} to create a coflow trace having the same characteristics as the Facebook trace from~\cite{Sincronia}. The trace contains 150 coflows, which are composed of 2086 total flows. In total, the Intra-pod traffic was 32.8 GB and the Inter-pod traffic 25.4 GB. We increase the workload by reducing inter-coflow arrival rates. As in~\cite{Chowdhury_2014,Chowdhury_2015}, we group coflows into \emph{short} if the longest flow of a coflow has less than 5MB. A coflow is classified as \emph{narrow} if it has less than 50 flows leading to four categories: Short and Narrow (SN), Long and Narrow (LN), Short and Wide (SW) and Long and Wide (LW).

\noindent
\textbf{Network Layer Load Balancing:} We compare Equal-Cost Multi-path (ECMP) and HULA~\cite{katta2016hula}. HULA is a flowlet-based load-balancing scheme that forwards flowlets over the least congested path. Fowlet gap is set to 500$\mu$s and probing interval is set to 200$\mu$s~\cite{katta2016hula}.

\noindent
\textbf{Transport Protocol and Queue:} We use  DCTP \cite{Alizadeh:2010}  with standard retransmission time-out of 3 RTTs and an RTO of 200us as in \cite{alcoz2020sp}. As baseline (deRED), we use 8 strict priority (SP) queues, each one holds  \cite{DiffservQueue} max. 500 packets. Each physical queue contains a single virtual RED queue ($min\_th=200$, $max\_th=400$) that starts marking ECN at $min\_th=200$ with a given probability and the scheduler maps flows to queues given by the Diffserv field. When using \textit{pCoflow}, the single queue has 8 priority bands of aggregated size. Each priority band starts marking packets at $min\_th=200$ per band. %

\begin{figure*}[!htb]
     \begin {minipage}[t]{0.32\textwidth}
     \centering
     \includegraphics[width=.9\linewidth]{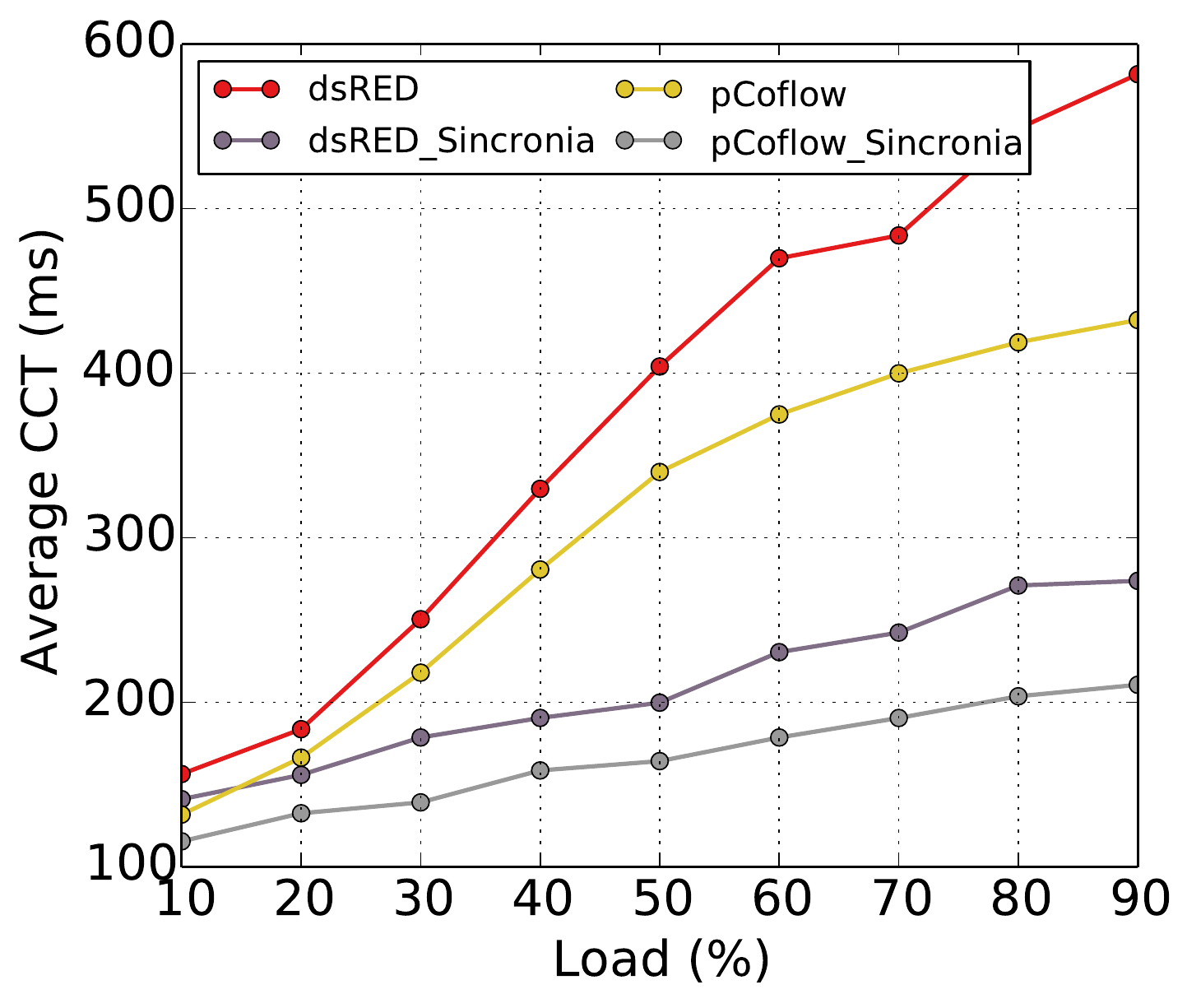}\par\caption{Average CCT for BigSwitch}  \label{fig:FIG3}
   \end{minipage}
   \begin{minipage}[t]{0.32\textwidth}
     \centering
     \includegraphics[width=.9\linewidth]{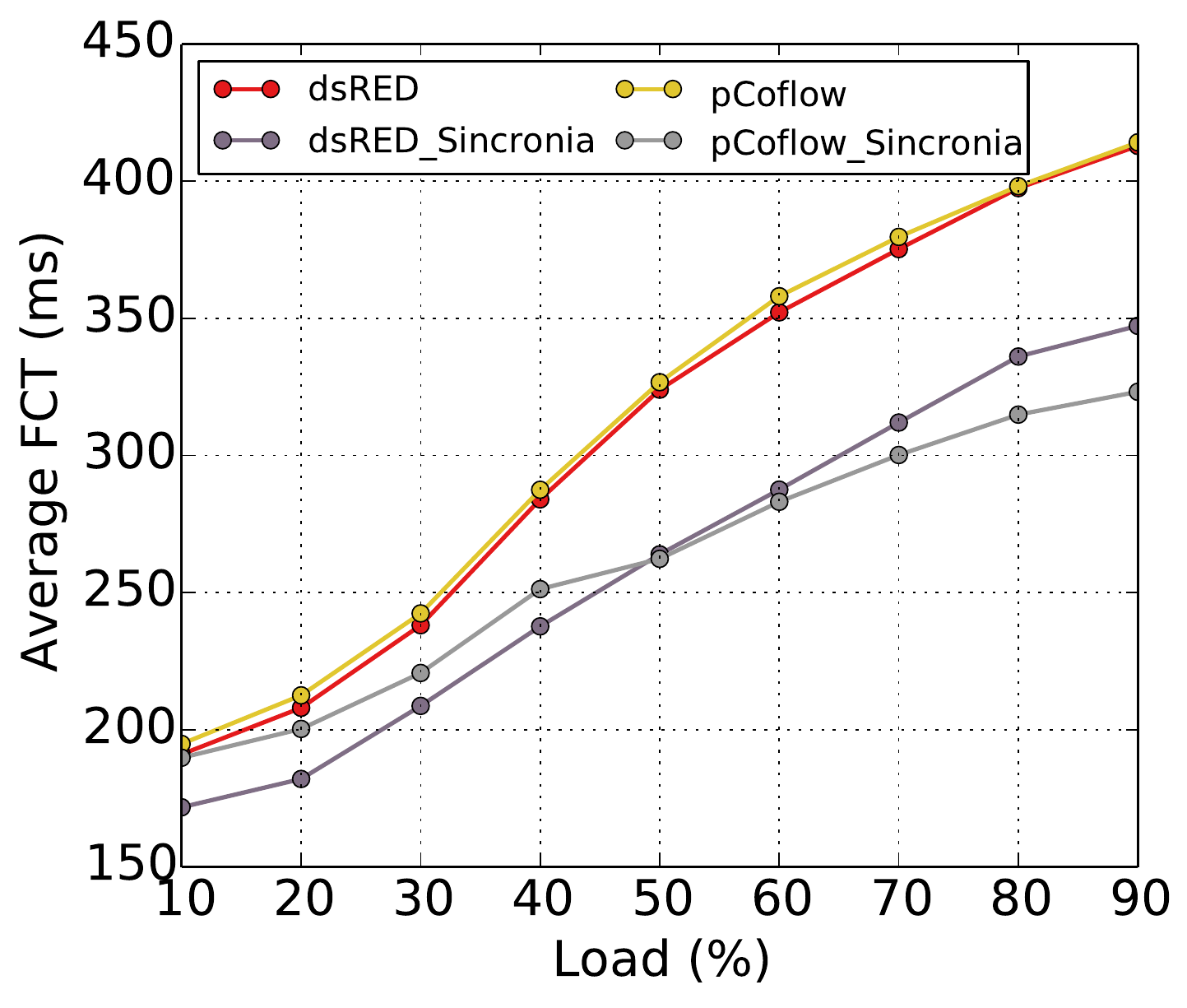}\par\caption{Average FCT for BigSwitch} \label{fig:FIG4}
   \end{minipage} \hfill
    \begin {minipage}[t]{0.32\textwidth}
     \centering
	\includegraphics[width=.9\linewidth]{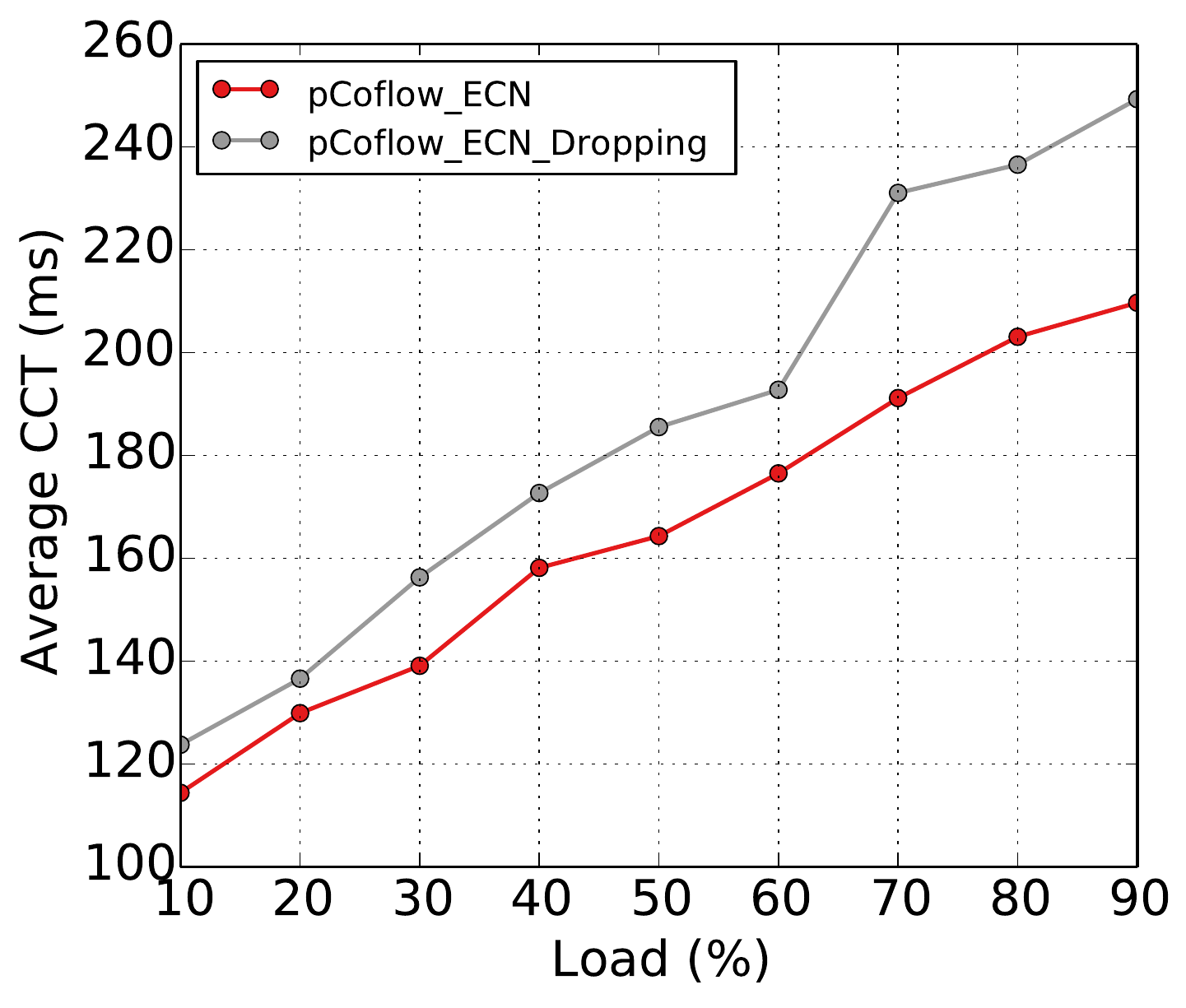}\par\caption{pCoflow Queue ECN vs ECN-Dropping} \label{fig:ECN_DROP}
   \end{minipage}
\end{figure*}

\begin{figure*}[!htb]
\begin{minipage}[t]{0.32\textwidth}
     \centering
     \includegraphics[width=.9\linewidth]{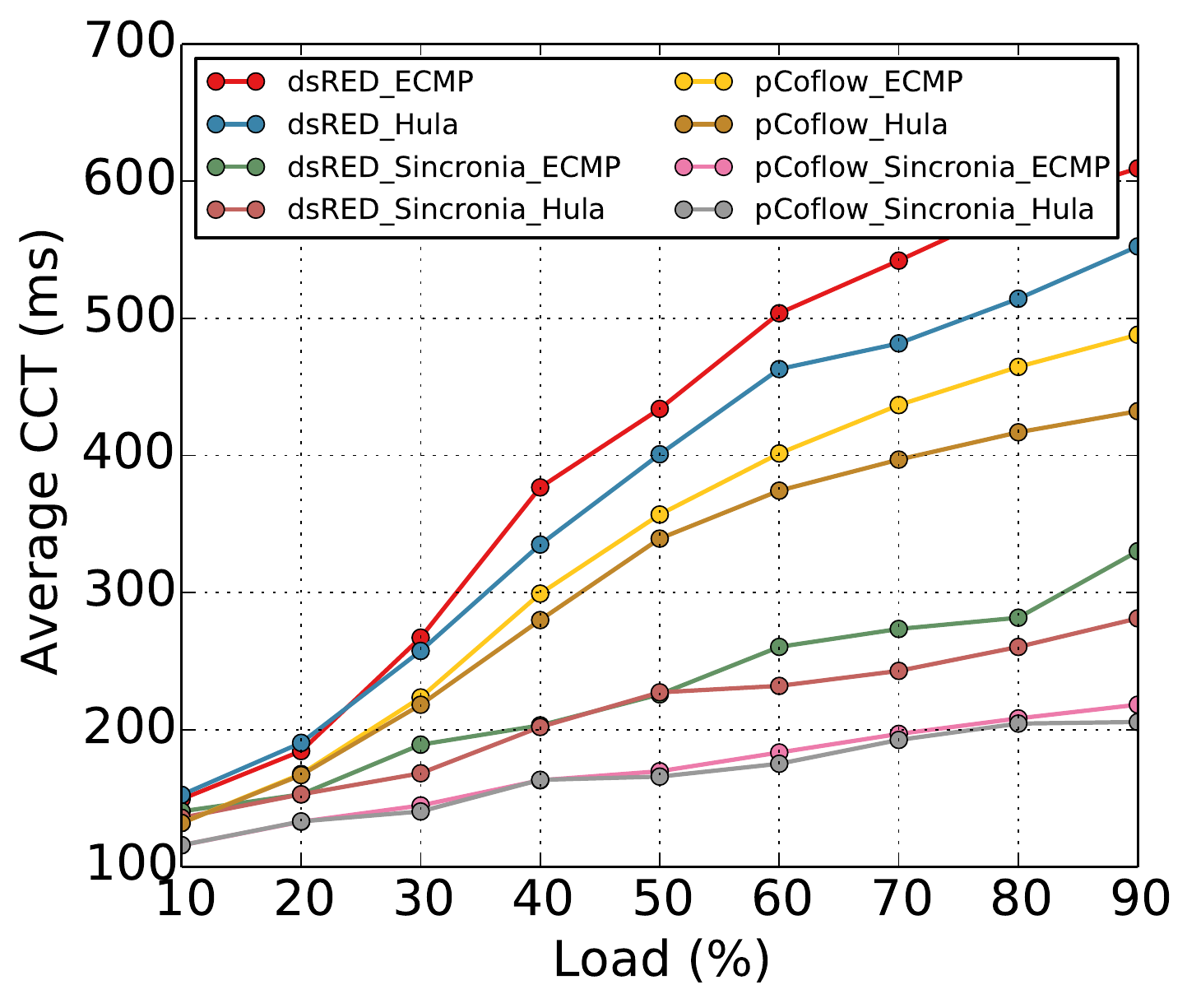}\par\caption{Average CCT for fat-tree topology} \label{fig:FIG1} 
   \end{minipage} \hfill
   \begin {minipage}[t]{0.32\textwidth}
     \centering
     \includegraphics[width=.9\linewidth]{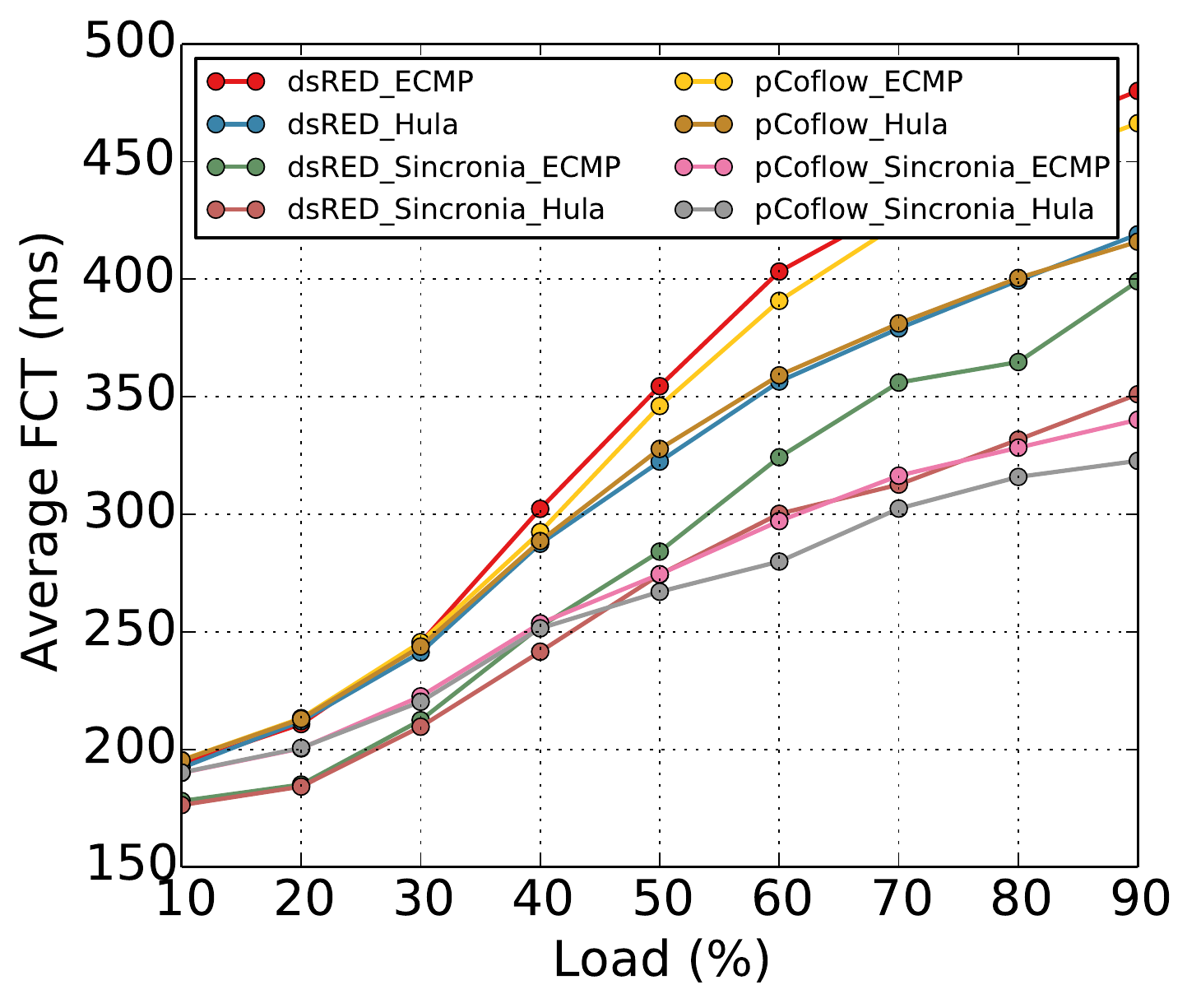}\par\caption{Average FCT for fat-tree topology} \label{fig:FIG2}
   \end{minipage} \hfill
   \begin {minipage}[t]{0.32\textwidth}
     \centering
     \includegraphics[width=.9\linewidth]{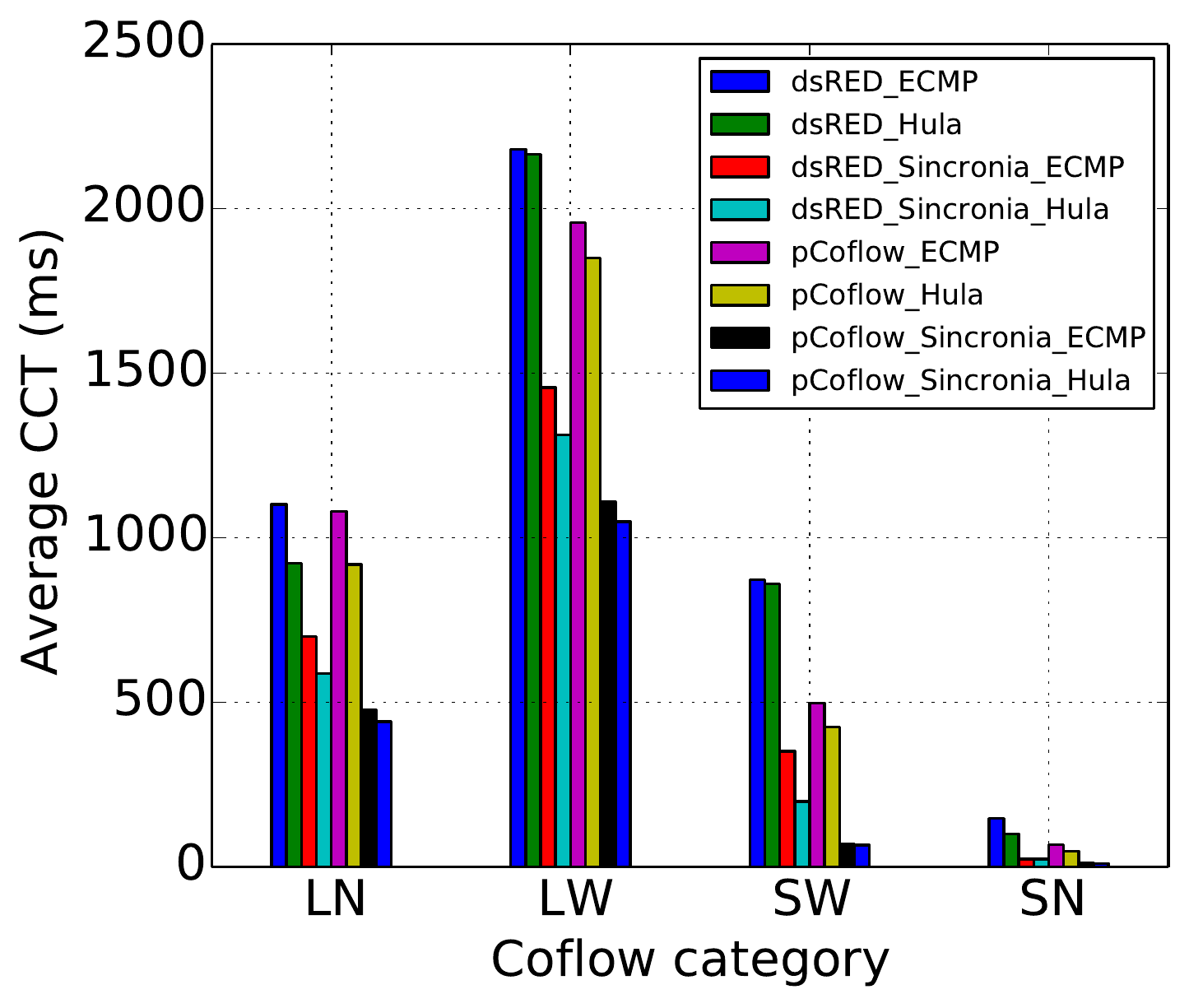}\par\caption{Average CCT for coflows sorted by category for 90\% load} \label{fig:FIG5} 
   \end{minipage} \hfill

\end{figure*}

\textbf{Results for BigSwitch:} The first question we try to answer for \textit{pCoflow} is, if it is better to drop the packets once the maximum number of packets per priority band is exceeded or allow to borrow space from lower priority bands? Figure \ref{fig:ECN_DROP} compares \textit{pCoflow\_Drop}, which drops packets once the limit for a band is reached (500) with \textit{pCoflow\_ECN} which adaptively adjusts queue bands, when using Sincronia for priority ordering. Dropping packets once the threshold is exceeded avoids coflow starvation by not allowing packets from other priorities to take queue space reserved for other priorities.  On the other hand, allowing coflows to temporary exceed their reserved queue space enables flows to steadily reduce their sending rate. Although this decision may temporarily lead to coflow starvation, we note that coflows can only take more space in the queue whenever there is space left from other coflows. In Figure \ref{fig:ECN_DROP}), we observe that allowing coflows to temporally exceed the priority band limit of 500 packets, we can reduce the overall CCT. This is due to DCTCP which reacts upon the ECN marking. All remaining experiments for \textit{pCoflow} are performed with adaptive queues and ECN marking. Figure \ref{fig:FIG3} and Figure \ref{fig:FIG4} show the average CCT and FCT for different combinations of load-balancing with and without Sincronia ordering.  \textit{pCoflow} improves both upon Sincronia and when not using Sincronia. When not using Sincronia, the benefits of \textit{pCoflow} are attributed to the adaptive queue size. \textit{pCoflow} improves upon  multi-level dsRED queues when using Sincronia since we avoid reordering and leverage the full capacity of the queue. At higher load, the benefits of \textit{pCoflow} are more pronounced, where the gap between our approach and the vanilla SP multi-level dsRED queue is in the range of 15-25\%. This might stem from the fact that at high loads the traffic is more unstable, leading to more changes in priorities and therefore more reordering. However, as Figure \ref{fig:FIG4} shows, by changing the insertion order of the packets, this can lead to an increase in the tailing of some flows and therefore leading to an increase of the FCT compared to the multi-level dsRED queue for low loads. For network loads higher than about 70\%, \textit{pCoflow} also reduces FCT compared to  other approaches. %\todo{Update values when replotting}

\textbf{Fattree - Summary:} Figure \ref{fig:FIG1} and Figure \ref{fig:FIG2} show the average CCT and FCT for the Fattree topology. The lowest CCT is achieved by \textit{pCoflow} when used with HULA (see Figure \ref{fig:FIG1}). When using Sincronia, the difference between ECMP and HULA is not significant since HULA probe packets are used to identify least congested paths. Consequently, we map them to  the highest priority queue or band. This can lead to a situation where low priority packets being forwarded to a more congested path. Moreover, Facebook traffic is characterized by having a one-to-many communication pattern, where a single node receives data from different nodes in the network. At low loads, the bottleneck might be located mainly on the links between the ToR and the server and therefore load-balancing plays a minor role. On the other hand, we can see that when we do not use Sincronia, the effect of load-balancing is more pronounced due to the congestion-aware load-balancing by HULA. When using Sincronia, \textit{pCoflow} combined with HULA achieves a CCT reduction up to 27\% compared to the fixed dsRED multi-level queuing when used with HULA. On the other hand, \textit{pCoflow}  can reduce CCT by 34\% compared to dsRED multi-level queues when used with Sincronia and ECMP.  Figure \ref{fig:FIG5} analyzes the CCT for the 90\% load case  for each coflow category. As expected, long and wide  (LW) coflows contribute to the highest CCT. This is because they are the coflows that transport the largest data volume. In addition, Sincronia benefits small coflows, as they have a higher probability to be assigned to a higher priority band. Surprisingly, the load-balancing scheme plays a less important role for large flows than expected, which may be caused by the unawareness of HULA of coflow properties. This leaves room for an integrated design with \textit{pCoflow}.

%% file: background_motivation.tex
There is extensive work related to scheduling coflows in data centers. 
Most of these schemes including \cite{Sincronia,Chowdhury_2014,zhao2015rapier,luo2015minimizing,chowdhury2011managing} rely on prior knowledge about coflows (e.g. flow sizes, server pairs). Coflow scheduling methods can be divided into two groups: \textit{distributed schedulers} and \textit{centralized schedulers}. Distributed schemes including \cite{luo2015minimizing,susanto2016stream,dogar2014decentralized} are executed on each host where coflows are scheduled and sorted locally. On the contrary, centralized schemes such as \cite{zhao2015rapier,Chowdhury_2014,li2016efficient,chowdhury2011managing} rely on a central controller to order coflows. Indeed, having a global view enables better scheduling decisions \cite{wang2018survey} while facing a a large control overhead and, therefore, scalability is an issue. Aalo \cite{Chowdhury_2015} uses priority queues to classify coflows according to the amount of data sent and does not need prior knowledge.
Sincronia \cite{Sincronia} overcomes the main centralized schedulers' problems by avoiding per-flow rate allocation. Sincronia achieves near-optimal average CCT and requires a transport layer proritizing flows according to coflow orderings. While most of the flows do not consider coflow routing, Rapier \cite{zhao2015rapier} and \cite{jahanjou2017asymptotically} demonstrate that combining scheduling and routing can lead to better performance.

%% file: conclusion.tex
We presented \textit{pCoflow}, an in-network support to coflow scheduling. 
Our work integrates state-of-the-art end-host coflow reordering approaches with in-network packet prioritization performed in the switch.
Our approach uses the PIFO scheduling abstraction to build a coflow aware packet scheduler which considers packet history during priority scheduling. \textit{pCoflow} therefore avoids excessive packet reordering that potentially lead to wasteful restransmissions. Our approach improves
 upon coflow completion time and benefits from flowlet-based load-balancing 
schemes such as HULA. As future work, we aim for an integrated design by defining extensions and proper interactions between \textit{pCoflow} and flowlet-based load-balancing schemes such as HULA.